\documentclass[aps,reprint,superscriptaddress,floatfix,showpacs,pra]{revtex4-1}
\usepackage{graphicx}
\usepackage{amsmath}
\usepackage{amssymb}
\usepackage{amsfonts}

\usepackage[unicode=true,breaklinks=true,colorlinks=true]{hyperref}
\hypersetup{citecolor=blue,urlcolor=blue}

\usepackage{newtxtext}
\usepackage{newtxmath}
\usepackage[all]{hypcap}
\usepackage{natbib}
\usepackage{multirow}
\usepackage{booktabs}
\usepackage{bbold}
\usepackage{bm}

\newcommand{\ket}[1]{|{#1}\rangle}

\newcommand{\braket}[2]{\langle{#1}|{#2}\rangle}

\begin{document}

\title{Quantum optimal control using phase-modulated driving fields}
\author{Jiazhao Tian}
\affiliation{School of Physics, International Joint Laboratory on Quantum Sensing and Quantum Metrology, Huazhong University of Science and Technology, Wuhan 430074 P. R. China}
\affiliation{State Key Laboratory of Precision Spectroscopy, East China Normal University, Shanghai, 200062, China}
\author{Haibin Liu}
\email{liuhb@hust.edu.cn}
\affiliation{School of Physics, International Joint Laboratory on Quantum Sensing and Quantum Metrology, Huazhong University of Science and Technology, Wuhan 430074 P. R. China}
\affiliation{State Key Laboratory of Precision Spectroscopy, East China Normal University, Shanghai, 200062, China}
\author{Yu Liu}
\author{Pengcheng Yang}
\affiliation{School of Physics, International Joint Laboratory on Quantum Sensing and Quantum Metrology, Huazhong University of Science and Technology, Wuhan 430074 P. R. China}
\author{Ralf Betzholz}
\affiliation{School of Physics, International Joint Laboratory on Quantum Sensing and Quantum Metrology, Huazhong University of Science and Technology, Wuhan 430074 P. R. China}
\author{Ressa S. Said}
\affiliation{Institute for Quantum Optics and Center for Integrated Quantum Science and Technology, Ulm University, D-89081 Ulm, Germany}
\author{Fedor Jelezko}
\affiliation{Institute for Quantum Optics and Center for Integrated Quantum Science and Technology, Ulm University, D-89081 Ulm, Germany}
\author{Jianming Cai}
\affiliation{School of Physics, International Joint Laboratory on Quantum Sensing and Quantum Metrology, Huazhong University of Science and Technology, Wuhan 430074 P. R. China}
\affiliation{State Key Laboratory of Precision Spectroscopy, East China Normal University, Shanghai, 200062, China}

\date{\today}

\begin{abstract}
  Quantum optimal control represents a powerful technique to enhance the performance of quantum experiments by engineering the controllable parameters of the Hamiltonian. However, the computational overhead for the necessary optimization of these control parameters drastically increases as their number grows. We devise a novel variant of a gradient-free optimal-control method by introducing the idea of phase-modulated driving fields, which allows us to find optimal control fields efficiently. We numerically evaluate its performance and demonstrate the advantages over standard Fourier-basis methods in controlling an ensemble of two-level systems showing an inhomogeneous broadening. The control fields optimized with the phase-modulated method provide an increased robustness against such ensemble inhomogeneities as well as control-field fluctuations and environmental noise, with one order of magnitude less of average search time. Robustness enhancement of single quantum gates is also achieved by the phase-modulated method. Under environmental noise, an XY-8 sequence constituted by optimized gates prolongs the coherence time by $50\%$ compared with standard rectangular pulses in our numerical simulations, showing the application potential of our phase-modulated method in improving the precision of signal detection in the field of quantum sensing.
\end{abstract}

\maketitle

\section{Introduction}\label{section1}
As the extension of optimal control theory in the quantum realm, quantum optimal control (QOC) seeks for the design of control fields that drive a quantum system to achieve certain tasks, such as state preparation~\cite{Rae_07_PRL,Car_06_PRL,Hoh_07_PRA}, wavefunction control~\cite{Tur_03_JPAMG,Wei_99_NL,Lan_14_PRL}, noise suppression~\cite{Gor_08_PRL,Don_19_PRA,Ji_16_PhyE}, or the implementation of quantum logic gates~\cite{Kha_05_JMR,Sch_14_NJP,Dol_14_NC,Kel_14_PRL,Mon_07_PRL,Amy_13_IEEE,Egg_14_PRL,Cho_14_PRL,Nai_11_PRL}. It covers a wide range of fields of physics, chemistry, biology, medicine, and their cross disciplines. For example, QOC has important applications in magnetic-resonance spectroscopy and imaging, quantum information processing, quantum simulation, and quantum sensing~\cite{Gla_15_EPJD}.

The most prevailing numerical methods used in QOC are gradient-based methods, including the GRAPE method~\cite{Kha_05_JMR,Jia_18_PRA,Jag_14_PRA,Fou_11_JMR,Tian_19_PRA} and the Krotov algorithm~\cite{Kro_95_xx,Mad_06_NM,Rei_12_JCP,Eit_11_PRA}. They are mathematically non-trivial and work in high-dimension control spaces~\cite{Sor_18_PRA}. On the other hand, gradient-free methods, e.g. the CRAB method and its variants~\cite{Can_11_PRA,Dor_11_PRL,Jac_14_PRL,Rac_15_PRA}, that transforms a functional optimization into a multi-variable function optimization, work in reduced control space, therefore are relatively simple to use and yet produce smooth control field. There are other methods however, including direct methods, which discretize the optimal control problem and use nonlinear optimization solvers~\cite{Li_09_JCP,Ste_19_PRA}, and synthetic methods, e.g., GOAT and GROUP~\cite{Mac_18_PRL,Sor_18_PRA}. Generally speaking, gradient-based methods are more involved but provide better local optimal results. They work well for problems with straightforward target functions and no constraint conditions~\cite{Rus_17_PTRSA}. By contrast, the gradient-free parameterization methods can handle problems with more diversified target functions more efficiently. They also allow for explicit restrictions by choosing a suitable search method and are favorable when the gradient information is hardly accessible.
Furthermore, the variety of easily available optimization solvers, make parameterization methods simple and flexible to use. However, the determination of a sufficient number of parameters in parametrized optimizations is crucial. Ideally, the number of parameters has to be increased until a convergence of the results is met. In practice, the balance between the improvement of the results and the increasing computational overhead has to be carefully maintained. Therefore, methods with sound optimal control results, less required parameters, and thereby shorter optimization time are highly demanded.

In this paper, we focus on the constrained optimization problem of the coherent control of a qubit ensemble with inhomogeneous broadening. We propose that a phase-modulated (PM) function basis is eminently suitable to construct the control field, since it shows the merit of finding favorable results in low-dimensional parameter spaces. Constrained by the same maximal field strength and the same maximal search resources, highly improved results are reached using one order of magnitude less of search time than the standard Fourier basis (SFB) used in the CRAB algorithm implementations~(see, e.g.,~\cite{Sch_14_NJP,Frank2017}). A phase-introduced version of SFB, referred to as SFB-P2, is proposed for a fairer comparison with the PM method in the view that both methods possess control fields in two directions in the interaction picture. SFB-P2 improves the results of SFB as the number of parameter increases, and gives comparable results with the $3$-parameters PM method using $20$ parameters and a search time one order of magnitude longer. Besides, among the same amount of total results with different initial values, PM possesses a much higher proportion of global optimal results than SFB-P2 with $20$ parameters, demonstrating a more stable capability to find the global optimal results with limited resource. These advantages can be attributed to the ability of PM to contain multiple frequency components in the control fields constructed using the PM basis with fewer parameters. In lower dimensional parameter spaces (i.e., $3$ parameters for PM and $4$ parameters for SFB-P2), these advantage become more prominent as the spectra of the ensemble inhomogeneity spread wider. Also, they hold in the presence of variations in the control fields and environment-induced dephasing, as the area bounded by the 90\% fidelity contour in the detuning and field-variation plane for PM is much larger than the one for SFB-P2 in both cases, with and without dephasing noise.
As a supplement, we also optimize the average fidelity of single qubit gates with inhomogeneous broadening using the PM method, and apply it to the XY-8 pulse sequence, where the simulated coherence time is prolonged by $50\%$ compared with the rectangular pulses of equal maximal pulse strength.
Combined with the randomization method, PM can be categorically classified into the CRAB-method family as a capable tool for solving problems including single spin systems under ambient noise, spin ensembles, and many-body systems.

The structure of the manuscript is as follows. In Sec.~\ref{sec_PM}, we describe the model of controlling an ensemble of two level systems with inhomogeneous broadening. In Sec.~\ref{sec_res}, a detailed comparison between the optimization performance using different choices of the function basis is demonstrated. The control field's robustness on unoptimized factors, including variations of field amplitude and environment-induced dephasing is shown in Sec.~\ref{sec_robust}. In addition, the optimization of single gate with inhomogeneous broadening and simulations of dynamical decoupling using our optimization results is presented in Sec.~\ref{sec_gate}. In Sec.~\ref{sec_con}, we conclude our work with a summary and the prospects of our proposal in QOC.

\section{Robust ensemble control model}\label{sec_PM}
Single-spin sensors, such as the nitrogen-vacancy (NV) center in diamond~\cite{Jel_06_PSSA,Doh_13_PR}, with their extraordinarily long coherence times at room temperature~\cite{Bal_09_NM,Mau_12_S} and their superb operability, e.g., simple initialization and readout by lasers~\cite{Gru_97_S}, are promising platforms for sensing applications with a high sensitivity and resolution~\cite{Gru_97_S,Jel_04_PRL,Maz_08_NL,Bal_08_NL,Gri_13_NP,Muel_14_NC,Shi_15_S,Sch_14_ARPC,Cai_14_NC}. Here, in scenarios such as wide-field sensing~\cite{Pha_11_NJP,Le_13_NL} or vector magnetometer~\cite{Sch_18_PRA,Ste_10_RoSI,Cle_18_APL}, ensembles of NV centers are suitably qualified for the performance of these tasks. However, due to the fact that each NV center may experience a slightly different local environment, an inhomogeneous broadening in the ensembles is inevitable. Similarly, for single spin measurements where the measurement cycle needs to be repeated multiple times, an inhomogeneous broadening in the temporal ensemble is present~\cite{Yang_16_RPP}. Furthermore, the control field amplitude may also differ between different NV centers in realistic scenarios. These two facts will thereby naturally degrade the fidelity of the control of NV ensembles. In order to overcome these difficulties, QOC methods have been widely adopted~\cite{Kha_05_JMR,Sch_14_NJP,Dol_14_NC,Dea_13_PRL,Zen_18_PRA}. Besides NV-center ensembles, other platforms utilizing ensembles are likewise facing similar difficulties, in the following we therefore focus on generic two-level quantum systems as the target of the QOC.

Driving a single two-level quantum system, with the energy eigenstates $\ket{\downarrow}$ and $\ket{\uparrow}$, from the initial state $\ket{\psi_0}=\ket{\downarrow}$ to the target state $\ket{\psi_g}=\ket{\uparrow}$ in a given time $T$ is, in principle, a simple task that only requires the application of a so-called $\pi$ pulse. However, due to the aforementioned inhomogeneity present in ensembles, $\pi$ pulses can only achieve high fidelities for a narrow band of the two-level systems, while leaving the majority of the ensemble poorly controlled ~\cite{bar_2015_thesis}. Therefore, the optimization problem we are facing here is the minimization of the harmful inhomogeneity effects by engineering an appropriate control field that achieves the best possible overall fidelity for the entire ensemble.

In the lab frame, the Hamiltonian that describes a single two-level system of the ensemble under a time-dependent control field $g(t)$ can be written as ($\hbar=1$)
\begin{equation}
  \label{totalH}
  H_\text{lab}(t) = \frac{\omega_0+\delta}{2}\sigma_z+g(t)\sigma_x,
\end{equation}
where $\omega_0$ is the energy splitting of the unperturbed spin, $\delta$ is the detuning caused by the inhomogeneity, and $\sigma_\kappa$ represents Pauli operators, for $\kappa=x,y,z$.

In order to describe the inhomogeneity of the two-level-system ensemble, here, we assume that the detunings $\delta$ follow a Gaussian distribution with zero mean value and a standard deviation $\sigma$~\cite{Nob_15_PRL,Vag_03_PRB}. The probability density $p(\delta)$ of the detuning distribution we consider thereby has the form
\begin{equation}
  \label{distrib}
  p(\delta)=\frac{1}{\sqrt{2\pi}\sigma}e^{-\frac{\delta^2}{2\sigma^2}}.
\end{equation}
In the following, we use the full width at half maximum (FWHM) $W=2\sqrt{2\ln{2}}\sigma$ to quantify the broadness of the distribution. Furthermore, we suppose that the spins which form the two-level systems of the ensemble are spatially sparse, such that their direct interactions can be neglected, which is a valid assumption for an ensemble of NV centers, whose direct dipole-dipole interaction can usually be discarded~\cite{Ben_13_PRL}, due to its rapid decrease with distance.

The performance of the ensemble control field is measured by the expectation value of $f (T,\delta)$ for a given probability density $p(\delta)$ of the detunings, which has the form
\begin{equation}
  \label{aveF}
  F=\int d \delta \, p(\delta)f (T,\delta),
\end{equation}
where $f(T,\delta)=\vert \braket{\psi_{g}}{\psi(T,\delta)}\vert^2$ is the fidelity of a final state $\ket{\psi(T,\delta)}$ and the target state for a given control time $T$ and detuning $\delta$. The target of the optimization task is then to maximize the value of $F$ or to minimize the value of the ensemble infidelity $1-F$~\cite{Can_11_PRA}.

\section{Phase-modulated control optimization}\label{sec_res}
\subsection{Method}
In order to maximize the value of $F$, we optimize the parameters that construct the control field $g(t)$ which drives the transition between the two energy eigenstates. Since $F$ is computationally expensive to directly evaluate, we take a small quantity ($M$) of evenly-spaced discrete detunings $\delta_k$, for $k=1,...,M$, to represent the whole Gaussian distribution~\cite{Nob_15_PRL}. The objective function of the optimization process is then written as
\begin{equation}
  \label{aveFd}
  F_\text{obj}=\mathcal{N}\sum_{k=1}^{M}p(\delta_k)f(\delta_k),
\end{equation}
with the normalization constant $\mathcal{N}=[\sum_{k=1}^M p(\delta_k)]^{-1}$.

In detail, we use $\delta_k\in[-W,W]$, covering around 98\% of the distribution's area, and $M=15$ to carry out the optimization process using a constrained nonlinear optimization algorithm (Nelder-Mead method) as our optimization solver. In order to evaluate the numerical performance of the optimization methods, we quantify the search speed by the number of objective function evaluations, which we represent by $n_f$.

Under the same constraints, different choices of the function basis for the control field $g(t)$ show distinct optimization efficiencies. In the widely used SFB the control field has the form
\begin{equation}
  \label{crab}
  g_\text{sfb}(t)=\sum_{j=1}^{N}{ a_j}\cos({\omega_j }t+{\phi_j})\cos(\omega_0t),
\end{equation}
where $N$ is the number of frequency components, while $a_j$, $\omega_j$ and $\phi_j$ are the parameters that have to be optimized. It can be seen that the phases $\phi_j$ are time independent, thus, depending on the number $N$, the control function $g_\text{s}(t)$ can be represented by only several discrete frequencies in Fourier space, which can be rather coarse.
\begin{figure*}[ht]
  \centering
  \includegraphics[width=17cm]{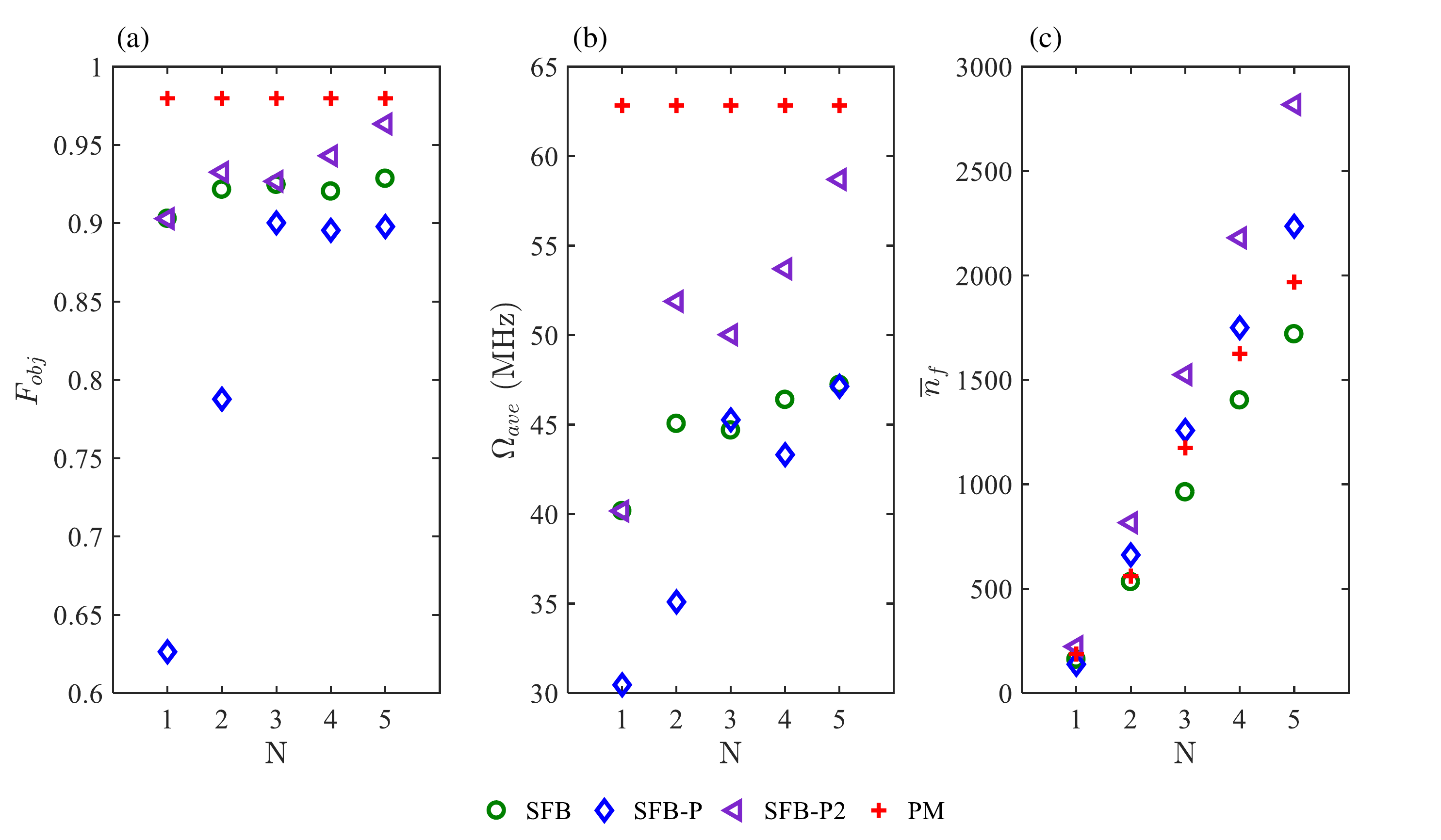}
  \caption{An comparison of optimal results of SFB, SFB-P, SFB-P2 and PM under the condition $T=100$ ns, $W/2\pi=10$ MHz, and amplitude constraint $\Omega_{\text{max}}/2\pi=10$ MHz. \textbf{(a)}: Optimal value of $F_{\text{obj}}$ as a function of $N$. \textbf{(b)}: Average field amplitude of the optimal fields as a function of $N$. \textbf{(c)}: Average number of function evaluations over $120$ optimal results as a function of $N$. Under the same constraints, the PM method reaches the best $F_{\text{obj}}$ using the least number of parameters, showing the efficiency of PM method on the premise of limited optimization resource.\label{Fig_FWT}}
\end{figure*}
In contrast to the SFB, inspired by the recently developed Floquet coherent controlling techniques, see, e.g.,~\cite{Rus_17_JSMTE,Shu_18_PRL,Sai_19_LP,Ber_20_ae}, we introduce a temporal modulation in the phases of the driving field according to
\begin{equation}
  \label{pm}
  g_\text{pm}(t)=\sum_{j=1} ^{N}a_j\cos\left[\omega_0t+\frac{{b_j}}{{\nu_j}}\sin({\nu_j}t)\right],
\end{equation}
with $a_j$, $b_j$ and $\nu_j$ being the optimization parameters. For a better demonstration of the increased frequency involvement in such a scheme, the above equation can be expanded as the Fourier series according to the Jacobi-Anger identity
\begin{equation}
  \label{pmex}
  g_\text{pm}(t)=\sum_{j=1} ^{N}a_j\sum_{l=-\infty}^{\infty}J_{l}\left(\frac{b_{j}}{\nu_{j}}\right)\cos\left[ (\omega_{0}+l\nu_{j})t\right],
\end{equation}
where $J_{l}(x)$ denotes the Bessel function of the $l$th order. It can be seen that the newly-introduced phase modulation enriches the complexity in Fourier space with an infinite number of frequency sidebands whose weight decreases with the distance from the central frequency $\omega_{0}$. As we will see, this feature greatly benefits QOC by making it possible to cover and exploit more frequencies with fewer parameters.

We note that both the SFB and the PM method only involve $\sigma_x$ components in the lab frame, however, in an interaction picture that use for the actual optimization, only the PM method will possess both $\sigma_x$ and $\sigma_y$ components; see Hamiltonians \eqref{Hs} and \eqref{Hp}. In order to nevertheless compare the efficiency of the two methods fairly, we further construct two forms of control fields by adding a phase term to the original SFB basis, thus both $\sigma_x$ and $\sigma_y$ components will be present in the interaction picture; see Hamiltonians~\eqref{Hsp} and~\eqref{Hsp2}. The first is denoted as SFB-P and has the form
\begin{equation}
  \label{SFB-P}
  g_\text{sfb-p}(t)=\sum_{j=1} ^{N}a_j\cos\left[\omega_0t+\omega_j t+\phi_j\right],
\end{equation}
which can be seen as a counterpart of PM and has the optimization parameters $a_j$, $\omega_j$,and $\phi_j$. The other is denoted by SFB-P2, which has a form similar to SFB but one more phase term $\varphi_j$, namely
\begin{equation}
  \label{SFB-P2}
  g_\text{sfb-p2}(t)=\sum_{j=1}^{N}{ a_j}\cos({\omega_j }t+{\phi_j})\cos(\omega_0t+\varphi_j),
\end{equation}
with $a_j$, $\omega_j$, $\phi_j$, and $\varphi_j$ being the optimization parameters.

As we mentioned before, for the sake of convenience, we carry out the numerical optimization in the interaction picture with respect to the central-frequency free Hamiltonian $(\omega_0/2)\sigma_z$. Neglecting counter-rotating terms in a rotating-wave approximation, the transformed lab-frame Hamiltonian~\eqref{totalH} can be written in the form $H(t)=H_0+H_\kappa(t)$, with $\kappa=\text{sfb}$, $\text{sfb-p}$, $\text{sfb-p2}$, $\text{pm}$ for the different bases. Here, the drift Hamiltonian is given by
\begin{equation}
  H_0=\frac{\delta}{2}\sigma_z,
\end{equation}
and the time-dependent control Hamiltonians read
\begin{gather}
  \label{Hs}
  H_{\text{sfb}}(t)=\sum_{j=1}^{N}\frac{a_j}{2}\cos(\omega_jt+{\phi_j})\sigma_x,\\
  \label{Hsp}
  H_{\text{sfb-p}}(t)=\sum_{j=1}^{N}\frac{a_j}{2}\left[\cos(\omega_jt+{\phi_j})\sigma_x+\sin(\omega_jt+{\phi_j})\sigma_y\right],\\
  \label{Hsp2}
  H_{\text{sfb-p2}}(t)=\sum_{j=1}^{N}\frac{a_j}{2}\cos(\omega_jt+{\phi_j})\left[\cos(\varphi_j)\sigma_x+\sin(\varphi_j)\sigma_y\right],\\
  \label{Hp}
  H_{\text{pm}}(t)=\sum_{j=1}^N\frac{a_j}{2}\left\{\cos\left[\frac{b_{j}}{\nu_{j}}\sin(\nu_{j}t)\right]\sigma_{x}+\sin\left[\frac{b_{j}}{\nu_{j}}\sin(\nu_{j}t)\right]\sigma_{y}\right\}
\end{gather}
for the SFB, SFB-P, SFB-P2, and PM methods, respectively. As one sees, all but the control field of the SFB method have both $\sigma_x$ and $\sigma_y$ components in the interaction picture.

In practical application there are some constraints imposed on the control fields, that should be taken into account in the optimization. For example, it is impossible to produce arbitrarily strong driving fields. That is, the maximum amplitude of the control field in our optimization will be bounded according to $\max |g(t)|\leqslant \Omega_\text{max}$. Additionally, in practical applications high frequencies are usually not preferable in the applied control fields, thus all frequency parameters, i.e., $\omega_k$, $b_k$ and $\nu_k$, are constrained to the range $2\pi\times$[0, $5/T $], in terms of the evolution time $T$. The phase parameters naturally fulfill $\phi_k,\varphi_k\in[0, 2\pi)$.
Since a single run of the optimization may result in a local optimal result, i.e., a local minimum in the control-parameter landscape, we always perform 120 runs with different starting points. Then, after the optimal control field is obtained, we randomly generate $K=10^5$ detunings $\delta_l$ obeying the Gaussian distribution~\eqref{distrib} in order to approximate the ensemble fidelity by $F\approx\sum_{l=1}^{K}f(\delta_l)/K.$

\subsection{Comparison of the methods}
\begin{figure}
  \centering
  \includegraphics[width=8cm]{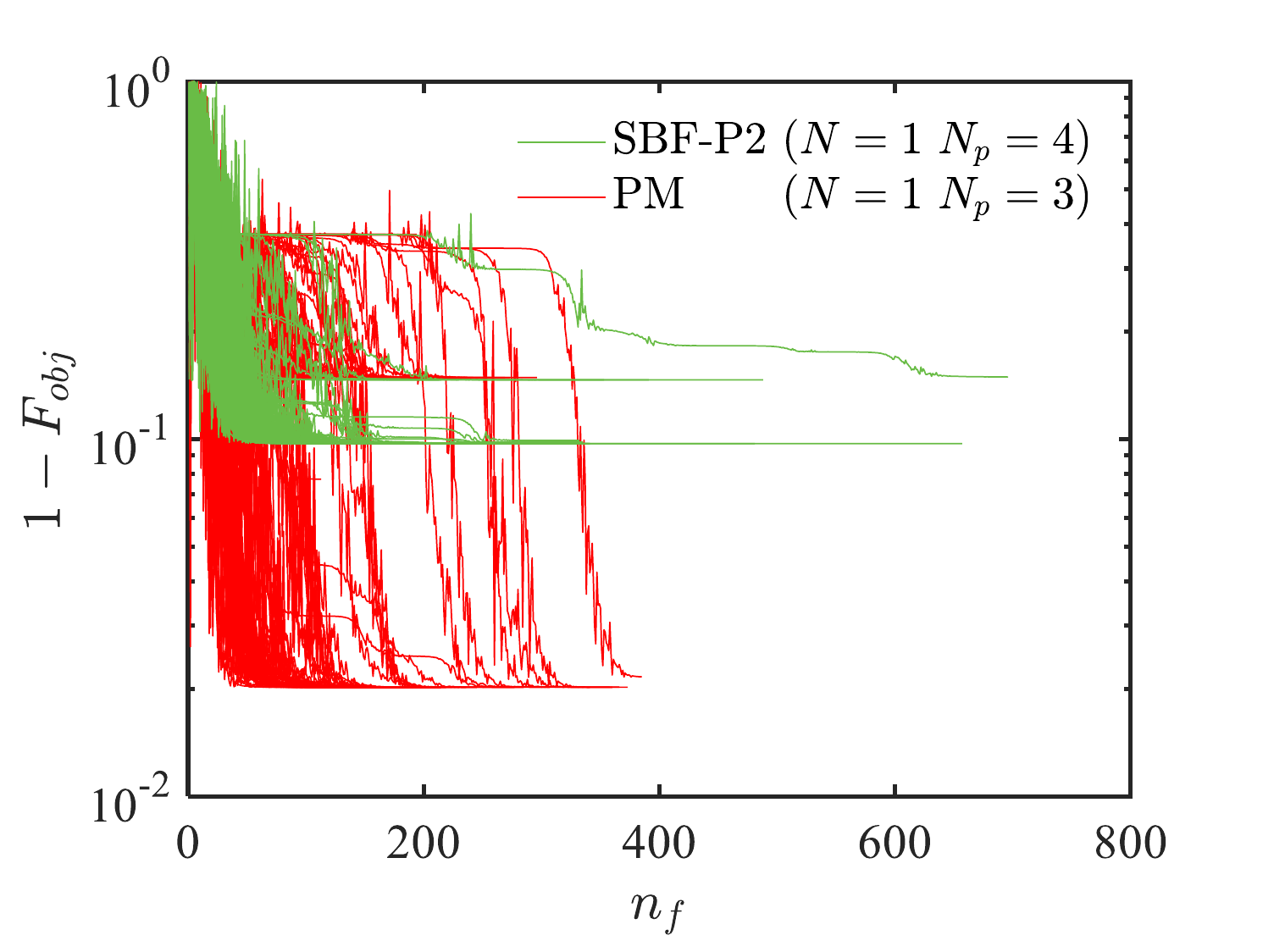}
  \caption{Infidelity of the final state as a function of the number of objective function evaluations ($n_f$). Red lines and green lines represent for the PM method with $3$ parameters and the SBF-P2 method with $4$ parameters respectively, both with 120 different random starting points. The parameters used here are $N=1$, $T=100$ ns, $W/2\pi=10$ MHz, and $\Omega_{\text{max}}/2\pi=10$ MHz. \label{fig-infidelity}}
\end{figure}
\begin{figure}
  \centering
  \includegraphics[width=8cm]{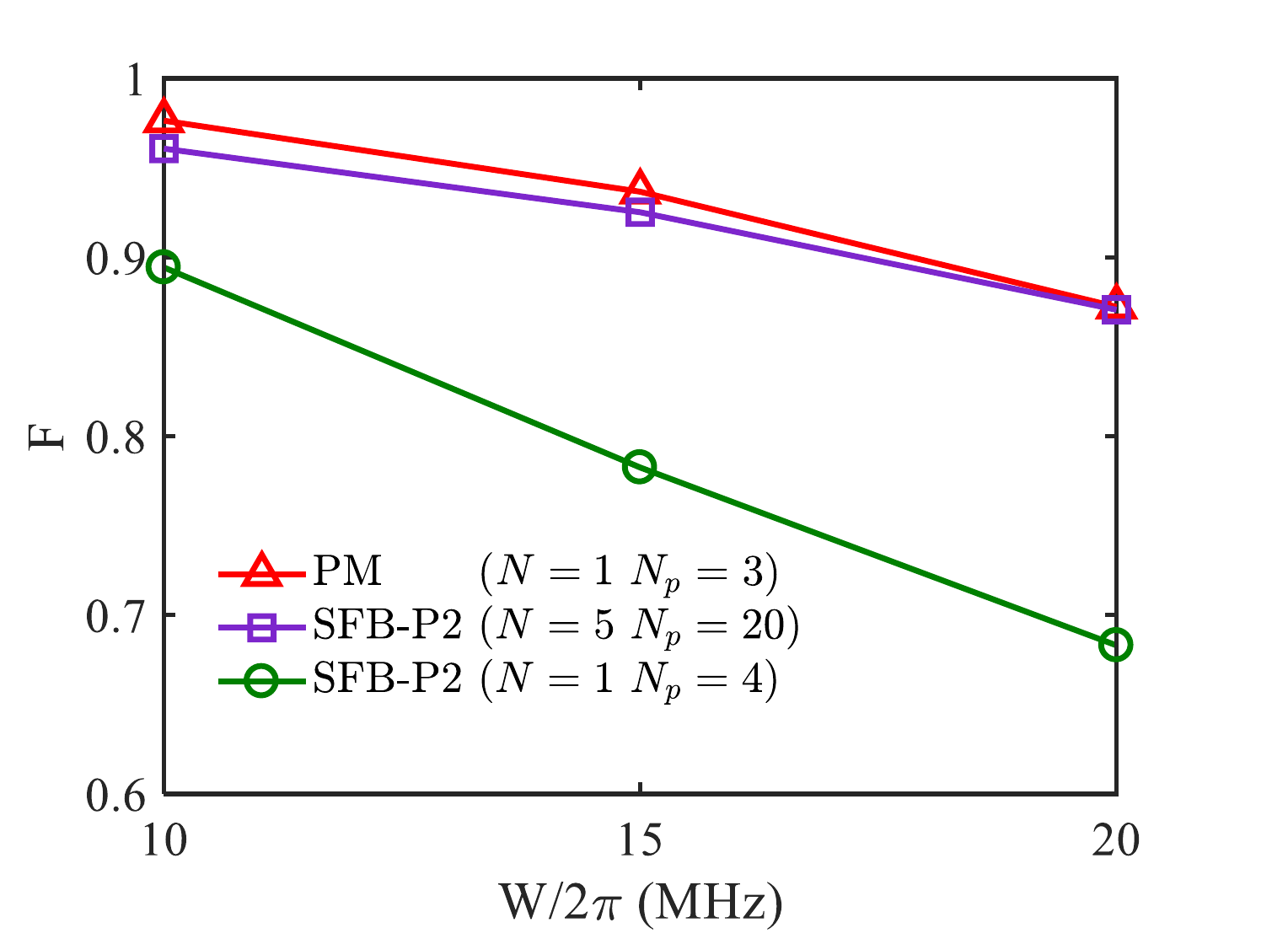}
  \caption{Optimal fidelity as a function of the FWHM ($W$) of the distribution of $\delta$. The red line with triangles is the result of the PM method with $N=1$, $N_p=3$. The purple line with square is the result of the SFB-P2 method with $N=5$, $N_p=20$. The green line with circles is the result of the SFB-P2 method with $N=1$, $N_p=4$.  Other parameters used here are $T=100$ ns, $W/2\pi=10$ MHz, and $\Omega_{\text{max}}/2\pi=10$ MHz. \label{fig-process}}
\end{figure}
The optimization results of PM, SFB, SFB-P, and SFB-P2 for $T=100$ ns and $W/2\pi=10$ MHz under the amplitude constraint $\Omega_{\text{max}}/2\pi=10$ MHz are shown in Fig.~\ref{Fig_FWT}. We make a comparison of these four methods in terms of three different aspects: the best value of the objective function $F_{\text{obj}}$, the average strength of the optimized control field $\Omega_{\text{ave}}=1/T\int \vert g(t)\vert dt$, and the number of function evaluations, which characterizes the search time, all shown as functions of $N$. Fig.~\ref{Fig_FWT}(a) and (b) show a basically positive correlation between $F_{\text{obj}}$ and $\Omega_{\text{ave}}$, which is in agreement with the common understanding that a stronger manipulation of a system mitigates the affects of noise better. Considering the optimization constraint in the maximum field amplitude $\Omega_{\text{max}}$, we may say that the PM method makes better use of the provided field power. In practice, although there are cases requiring a limited total power of the control field (e.g., in biology environments~\cite{Cao_20_PRap}), the maximal power is the primary limitation of the field generator~\cite{Nob_15_PRL, Bow_13_RSI}, which makes our proposed method reasonable. Fig.~\ref{Fig_FWT}(c) shows the rising trend of the average search time with the increase of $N$, which is a natural consequence of the expansion of the search space. Here, we use the default maximum number of function evaluations, which equals to $200\times N_p$, with $N_p$ the number of parameters. Increasing the maximum number of function evaluations could possibly improve the final results at the expense of a prolonged optimization time.

It is noteworthy to mention that, in general, the better efficiency of the PM method is guaranteed on the premise of a limited optimization resource, which is a common situation for most realistic optimization tasks. In Fig.~\ref{Fig_FWT}, the PM method shows its efficiency clearly using only $3$ parameters ($N=1$) and the shortest searching time while obtaining the largest value of $F_{\text{obj}}$. The optimal value of $F_{\text{obj}}$ given by SBF-P2 could merely catch up with PM using $20$ parameters ($N=5$). A comparison between SBF and SBF-P2 implies that introducing the phase term in a proper way can improve the capability of the SBF, as the number of parameters is increased. This could be attributed to the two-directional control field in the interaction picture that the phase term entails, although the inferior performance of SBF-P indicates that this is not always the case.

\begin{figure*}
  \centering\label{fig-spectrum}
  \includegraphics[width=17cm]{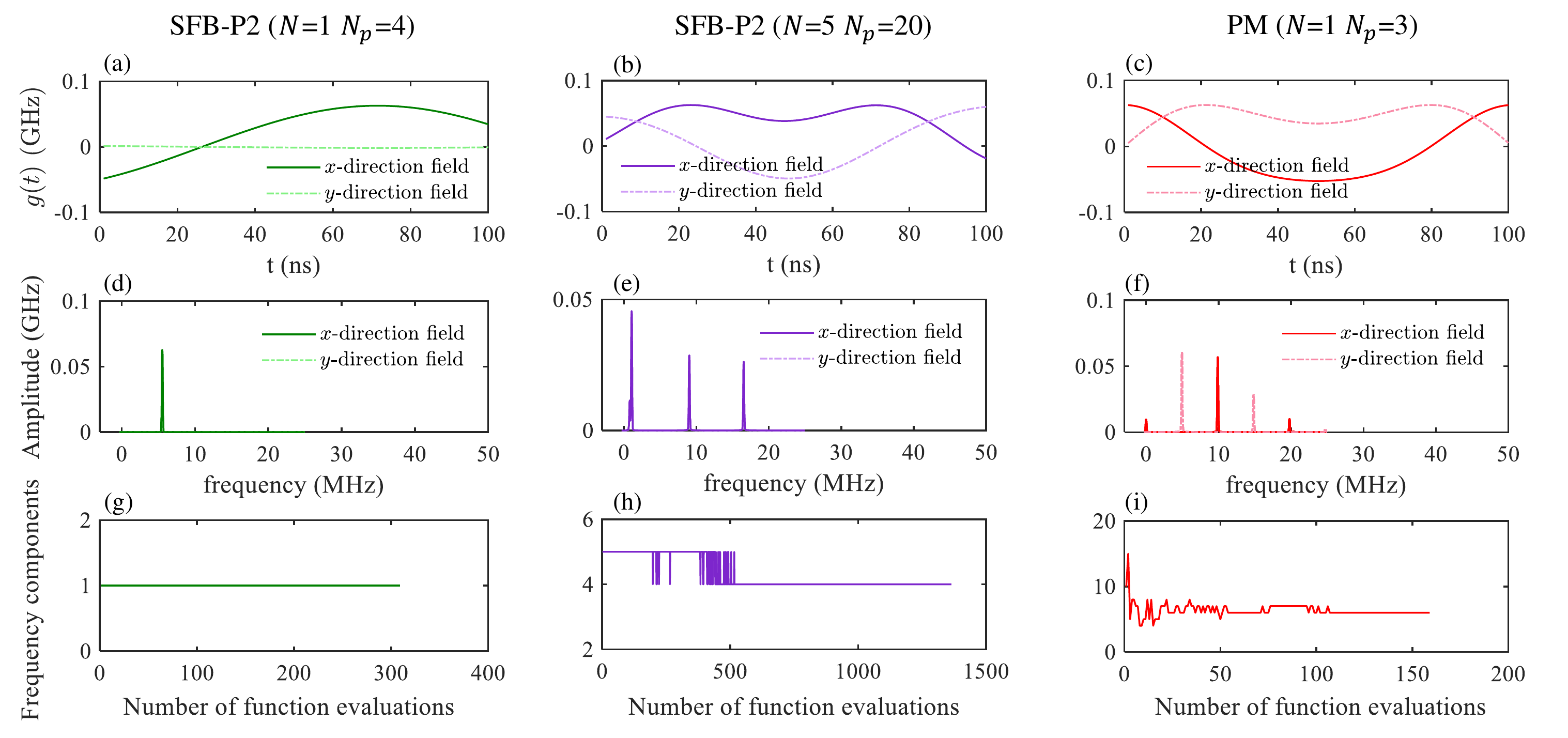}
  \caption{\textbf{(a-c)} Optimal control field in the interaction picture, given by the SFB-P2 ($N=1$, $N_p=4$), SFB-P2 ($N=5$, $N_p=20$) and PM method ($N=1$, $N_p=3$) respectively. Solid line: $x$-direction field, dotted dashed line: $y$-direction field. \textbf{(d-f)} The corresponding frequency spectrum of the control fields in \textbf{(a-c)} respectively. \textbf{(g-i)} The corresponding number of frequency components that amplitude $\geqslant 5$
  MHz during the optimization process.}
\end{figure*}
\begin{figure*}
  \centering\label{fig-landscape}
  \includegraphics[width=17cm]{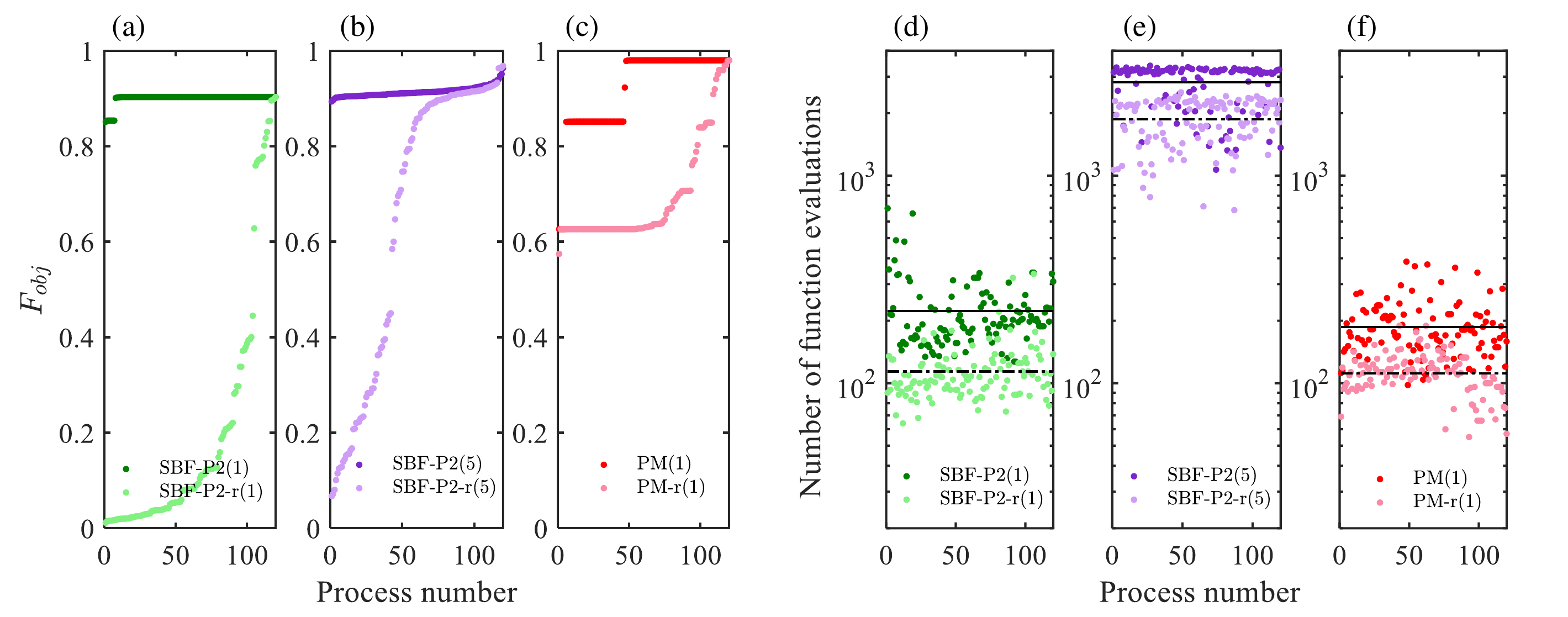}
  \caption{\textbf{(a-c)} Results of $120$ optimization process with different initial values, ranked by the value of $F_\text{obj}$. \textbf{(a)} The SFB-P2 method with $N=1$, $N_p=4$ and its parameter randomized version SFB-P2-r where $\omega_i$ are randomly taken from the range $2\pi\times$[0, $5/T $] and not being optimized. \textbf{(b)} The SFB-P2 method with $N=5$, $N_p=20$ and its parameter randomized version SFB-P2-r where $\omega_i$ are randomly taken from the range $2\pi\times$[0, $5/T $] and not being optimized. \textbf{(c)} The PM method with $N=1$, $N_p=3$ and its parameter randomized version PM(r) where $\nu_i$ are randomly taken from the range $2\pi\times$[0, $5/T $] and not being optimized. \textbf{(d-f)} Corresponding number of function evaluations (dot) and the average value (solid line for non-randomized methods and dashed line for randomized methods).}
\end{figure*}

\begin{figure*}
  \centering
  \includegraphics[width=17cm]{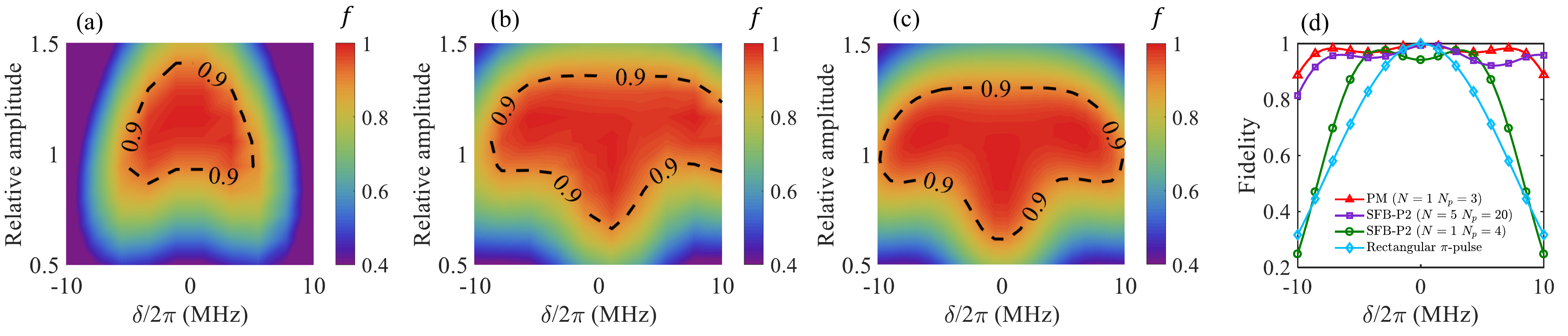}
  \caption{Fidelity of the final state under different values of the detuning and the relative variations of the control-field amplitude. \textbf{(a)} Results under a control field optimized using the SFB-P2 with $N=1$, $N_p=4$. \textbf{(b)} Results under a control field optimized using the SFB-P2 method with $N=5$, $N_p=20$. \textbf{(c)} Results under a control field optimized using the phase-modulated method with $N=1$, $N_p=3$. \textbf{(d)} A comparison of the three optimal methods and the rectangular $\pi$-pulse when there are no variations of the control amplitude. The ratio of areas of $f>0.9$ in \textbf{(a)} and \textbf{(c)} is approximately $1:1.98$. The markers in \textbf{(d)} represent the $M=15$ sample points used in the objective function during the optimization process. Other parameters used here are $T=100$ ns, $W/2\pi=10$ MHz, and $\Omega_{\text{max}}/2\pi=10$ MHz. The amplitude of the rectangular $\pi$-pulse is $2\pi\times 10$ MHz, with the pulse length of $50$ ns. \label{fig-amp}}
\end{figure*}

In order to have a more detailed look into the optimization process, we depict the trajectories of the optimization processes with red lines (PM method) and green lines (SBF-P2 method) in Fig.~\ref{fig-infidelity}, both with the same number of basis functions, namely $N=1$. It clearly shows the distinct advantages of PM method, i.e. much better results in relatively shorter searching time. This advantage becomes more prominent as the inhomogeneous broadening become wider, as Fig.~\ref{fig-process} displays.

In the following, using PM with $N=1$ and SBF-P2 with $N=1$ as well as $N=5$, we provide a further analysis based on our numerical optimization results. The optimal control fields in the interaction picture are demonstrated in Fig.~\ref{fig-spectrum}(a,b) for the SFB=P2 and Fig.~\ref{fig-spectrum}(c) for the PM method, respectively, and the frequency spectra of their control fields in the range $[0,50]$ MHz are shown in Fig.~\ref{fig-spectrum}(d-f). Fig.~\ref{fig-spectrum}(g-i) show the number of frequency components (counted when the amplitude is higher than 5 MHz) during the optimization process, revealing the potential of the PM method of covering more frequencies with fewer parameters. For SBF-P2, the frequency spectra are the same for the $x$- and $y$-directional fields, while for PM this is not the case, due to the property $J_{-l}(x)=(-1)^lJ_{l}(x)$, for $l\in \mathbb{N}$, of the Bessel functions.

Moreover, to explore the stability and convergence of the different methods, Fig.~\ref{fig-landscape}(a-c) presents $120$ results with each method ranked by the value of $F_{\text{obj}}$. For the PM method, more than half of these results reach the best possible value, showing that $120$ optimal processes are adequate for PM to obtain the global optimum. In contrast, the distribution of results of SBF-P2 with $N=5$ shows that $120$ processes are not sufficient to guarantee a global optimum. Fig.~\ref{fig-landscape}(d-f) exhibit the corresponding number of function evaluations for each process (dots) and their average value (solid and dashed-dotted lines), reflecting the search time for each optimization method.

As mentioned before, when equipped with a randomization, the PM method obviously belongs to the CRAB family, hence it enriches the toolbox of direct search optimization methods using truncated function bases. In Fig.~\ref{fig-landscape} the results with randomized frequencies (i.e., $\omega_j$ and $\nu_j$ for SBF-P2 and PM, respectively) are presented using light colors. While the average search times decreased, the best possible value, as well as the distribution of the optimization results, shows a disadvantage compared to the unrandomized case. Therefore, for the specific problem we discussed here, a randomization is not necessary. However, it might be found useful and even necessary for other cases.

\section{Robustness against unoptimized factors}\label{sec_robust}
\subsection{Imperfect control fields}
\begin{figure}
  \centering
  \includegraphics[width=8cm]{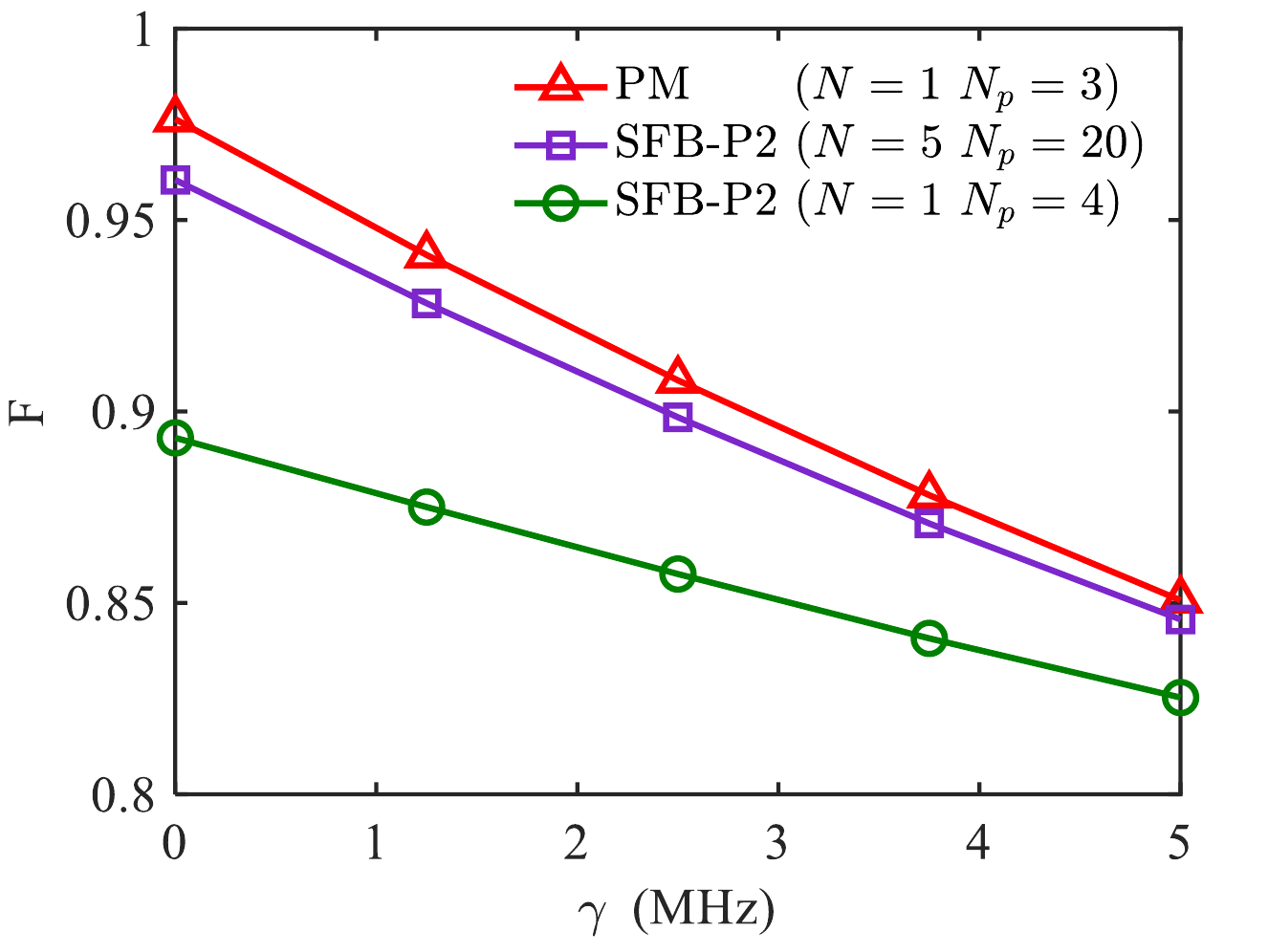}
  \caption{Average fidelity $F$ of the ensemble system as a function of the dephasing rate, using the same optimized control fields as in Fig.~\ref{fig-amp}. \label{figcoh_1}}
\end{figure}
\begin{figure*}
  \centering
  \includegraphics[width=17cm]{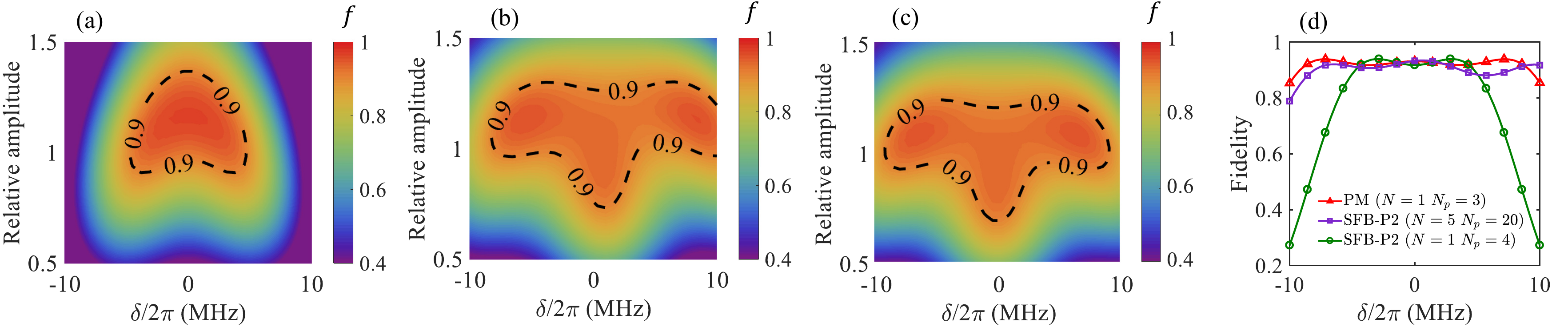}
  \caption{Fidelity of the final state with dephasing rate $\gamma=2$ MHz, using the same control fields and parameters as in Fig.~\ref{fig-amp}. \textbf{(a)} Results under a control field optimized using the SFB-P2 with $N=1$, $N_p=4$. \textbf{(b)} Results under a control field optimized using the SFB-P2 method with $N=5$, $N_p=20$. \textbf{(c)} Results under a control field optimized using the PM method with $N=1$, $N_p=3$. \textbf{(d)} A comparison of the three optimal methods when there are no variations of the control amplitude. The ratio of areas of $f>0.9$ in \textbf{(a)} and \textbf{(c)} is approximately $1:1.74$. The markers in \textbf{(d)} represent the $M=15$ sample points used in the objective function during the optimization process. \label{figcoh_2}}
\end{figure*}
In realistic experiments, variations of the control-field amplitude among different two-level systems in an ensemble are another obstacle which has to be overcome to achieve a high-fidelity control over the ensemble. To describe the influence of the control amplitude variations, we add a linear scaling factor $\alpha$ that amplifies or attenuates the control field amplitude. Thus, the Hamiltonian~\eqref{totalH} is rewritten as
\begin{equation}
  H= \frac{\omega_0+\delta}{2}\sigma_z+\alpha g(t)\sigma_x,
\end{equation}
where a time-independent scaling factor $\alpha$ is assumed for simplicity.

With the optimized parameters of each method, we obtain the fidelity of a single two-level system under different values of the detuning and the relative amplitude variation factor, as shown in Fig.~\ref{fig-amp}. Comparing Fig.~\ref{fig-amp}(a-b) [SFB-P2 method] with (c) [PM method], it is apparent that using a similar number of parameters the PM method shows an improved fidelity in a much wider range of the detuning and the control field variation (for $N=1$ with a $98\%$ increase of the area where $f>0.9$ compared to SFB-P2), i.e., again showing a stronger robustness, which the SFB-P2 need at least $20$ parameters to reach. A comparison of these methods when the relative amplitude equals to $1$ [i.e., a horizontal cut along (a-c)] is also presented in Fig.~\ref{fig-amp}(d). Here, we also show the fidelity for a simple $\pi$ pulse in the light blue line, viz., for a control Hamiltonian $H_\pi=(\pi/2T)\sigma_x$ in the interaction picture, which suggests all the optimized results, especially the one from PM method, show better robustness against inhomogeneities than the $\pi$ pulse.

\subsection{Effects of Noise}
In practice, the existence of environmental noise leads to the decoherence of quantum systems.
In spin systems, the longitudinal and the transverse relaxation are responsible for the loss of polarization along and perpendicular to the direction of the Pauli operator $\sigma_z$, respectively. Considering ensembles of NV centers that can be well described by two-level systems, in the electronic ground-state manifold the energy gap between the $\ket{m_s=0}$ and the $\ket{m_s=\pm1}$ states due to the zero-field splitting is roughly $2.8$~GHz~\cite{Jel_06_PSSA,Doh_13_PR}, which is much larger than the noise frequencies, which are usually on the order of a few MHz. This suppresses the longitudinal relaxation and, therefore, we only consider the perpendicular relaxation in the form of pure dephasing, with dephasing times ranging from hundreds of nanoseconds to a few microseconds~\cite{Sch_14_ARPC}.

Under such pure dephasing, the dynamics of a single two-level system density operator $\rho$ can be described by the Lindblad master equation~\cite{breuer_02_book,Yang_16_RPP}
\begin{equation}
  \frac{\partial}{\partial t}\rho=-i[H,\rho]+\frac{\gamma}{2}(\sigma_z \rho \sigma_z-\rho),
\end{equation}
where $\gamma$ is the reciprocal of the dephasing time $T^*_2$.

For simplicity, we assume that the dephasing rates of each two-level system are equal to the average dephasing rate of the ensemble. Since the final state evolved under noise is generally a mixed state, which is described by the density operator $\rho(T,\delta)$, the fidelity between the final state and the target state has to be written as
\begin{equation}
  f(\delta)=\left\langle \psi_{g}\right|\rho(T,\delta)\left|\psi_{g}\right\rangle.
\end{equation}

In Fig.~\ref{figcoh_1}, we show the average fidelity $F$ using control fields optimized under the PM and the SFB-P2 method, respectively, as a function of the dephasing rate $\gamma$. One can clearly see that, not surprisingly, the fidelity decreases as the dephasing rate becomes larger. However, under reasonable noise levels, e.g., up to 2 MHz for NV-center ensembles, the achieved fidelity is still sufficiently high to be applicable. Also, under dephasing noise, the PM results surpass the SBF-P2 method results, showing a better robustness against noise. Additionally, the robustness against detuning and control field variations under dephasing noise is depicted in Fig.~\ref{figcoh_2}. Compared to their non-dephasing counterparts in Fig.~\ref{fig-amp}, the $f>0.9$ area shrinks while maintaining similar profiles. The ratio of $f>0.9$ areas of PM and SFB-P2 ($N=1$) is $1.74:1$, approximately.

\section{Gate Synthesis and Dynamical Decoupling}\label{sec_gate}
\begin{figure}
  \centering
  \includegraphics[width=8cm]{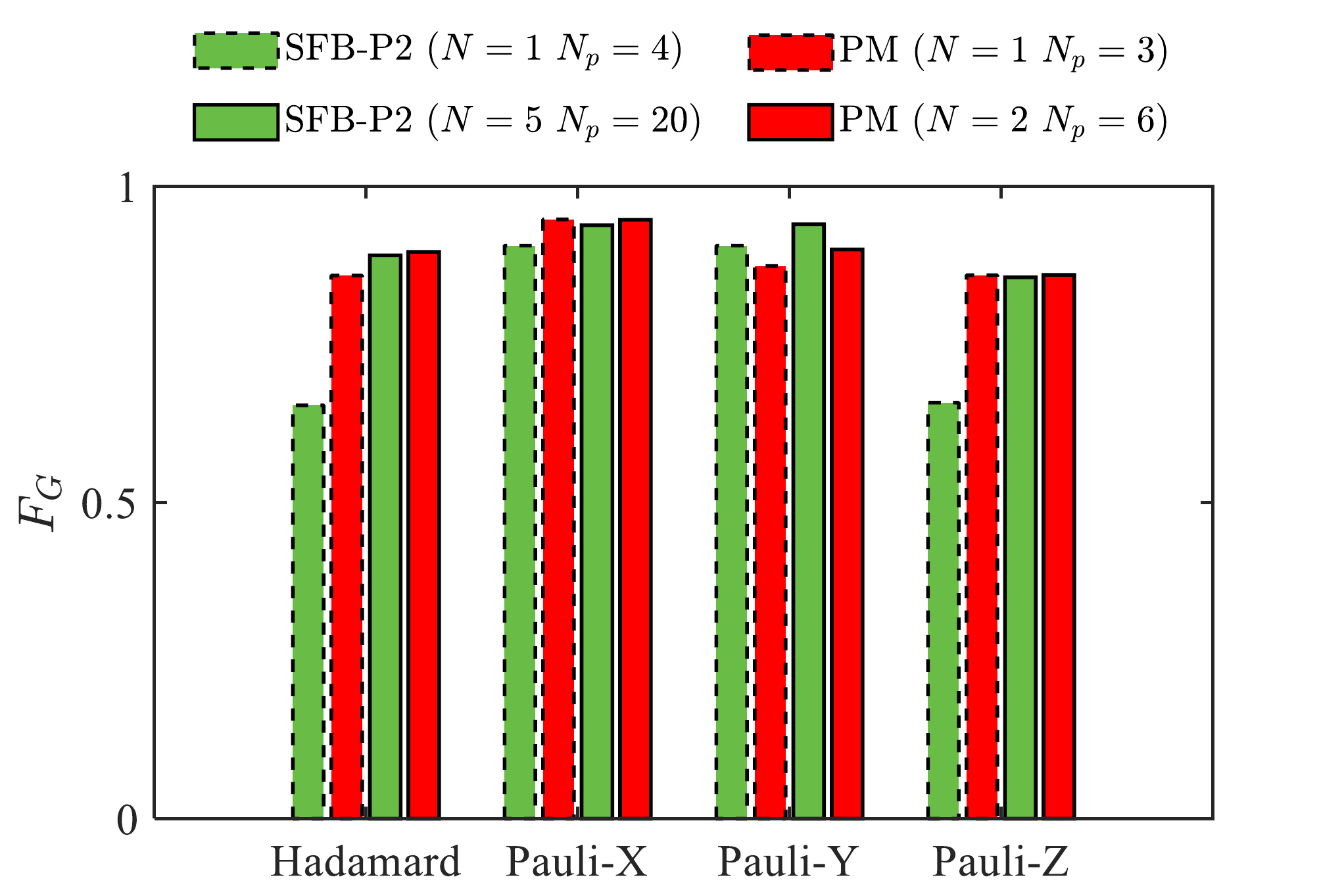}
  \caption{Optimal results of the robust gate fidelity given by PM ($N=1,2$) and SFB-P2 ($N=1,5$) methods, with the targeted Hadamard gate, Pauli-X gate, Pauli-Y gate and Pauli-Z gate respectively. Here we use $K=10^5$ random detunings $\delta_l$ obeying the Gaussian distribution~\eqref{distrib} with $W/2\pi=10$ MHz, to approximate the ensemble fidelity by $F_G\approx\sum_{l=1}^{K}f_g(\delta_l)/K$. Parameters used in the optimization process are $T=100$ ns, $W/2\pi=10$ MHz, and $\Omega_{\text{max}}/2\pi=10$ MHz. \label{fig9}}
\end{figure}
\begin{figure*}
  \centering
  \includegraphics[width=17cm]{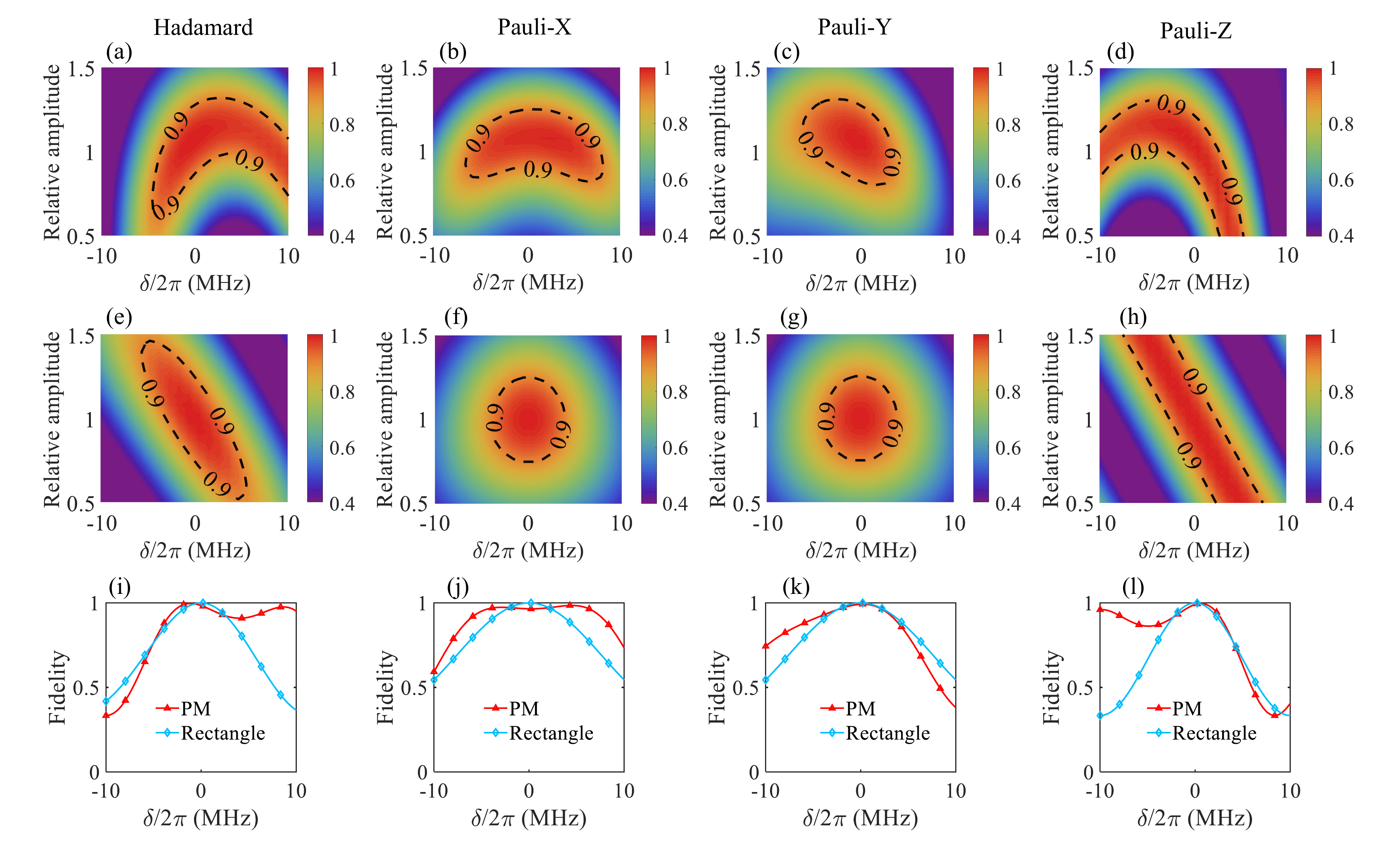}
  \caption{Gate fidelity under different values of the detuning and the relative variations of the control-field amplitude. \textbf{(a-d)} Results of the control fields optimized using the phase modulated method with $N=2$, with the target gate being the Hadamard gate, Pauli-X gate, Pauli-Y gate and Pauli-Z gate respectively. Parameters used in the optimization process are $T=100$ ns, $W/2\pi=10$ MHz, and $\Omega_{\text{max}}/2\pi=10$ MHz. \textbf{(e-h)} Results of the standard rectangular pulse with amplitude of $2\pi\times10$ MHz and pulse length $50$ ns, with the target gate being the Hadamard gate, Pauli-X gate, Pauli-Y gate and Pauli-Z gate respectively. \textbf{(i-l)} A comparison of optimized PM pulse and the rectangular pulse when there are no variations of the control amplitude. \label{fig10}}
\end{figure*}
\begin{figure}
  \centering
  \includegraphics[width=8cm]{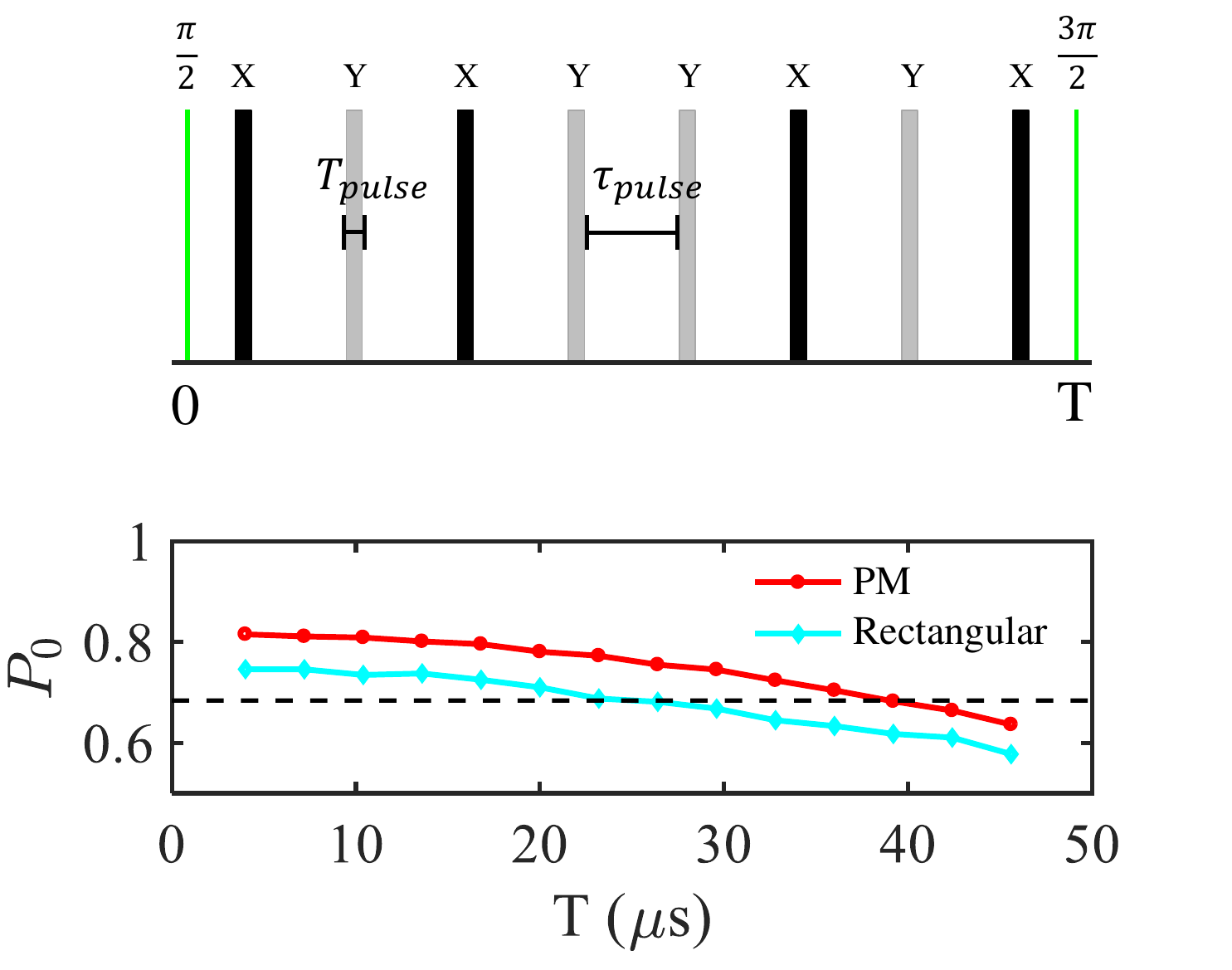}
  \caption{Simulation of the coherence time $T_2$ using one XY8 series. The initial state is taken as $\psi_0=\ket{0}$. \textbf{(a)} Schematic pulse location during one simulation precess with evolution time $T$. The slim green lines at the beginning and the end represent the initialization pulse rotating the state around X axis by $\pi/2$ and the readout pulse rotating the state around X axis by $3\pi/2$ respectively, both are presumed to be instantaneous. The black and gray sticks represent Pauli-X and Pauli-Y gate in the XY8 series respectively, with finite pulse length $T_{pulse}$. The pulse separation $\tau_{pulse}$ refers to the time period from the end of one pulse to the beginning of next pulse. The pulse number and pulse length is fixed for different evolution time $T$, and $T_2$ is recognized as the value of $T$ at which the population of $\ket{0}$ drops to $(1+1/e)/2$. \textbf{(b)} The population $P_0$ of $\ket{0}$ as a function of $T$. Each point averages the results of $1200$ evolutions. PM optimal control fields gives $T_2 \approx 39 \mu$s, improving the coherence time with rectangular pulse $T_2 \approx 24 \mu$s by half.\label{fig11}}
\end{figure}
The above optimization of state transitions can be easily extended to produce robust single qubit gates by replacing the state fidelity with the gate fidelity
\begin{equation}\label{fgate}
f_g(\delta)=\frac{1}{2}+\frac{1}{3} \sum_{\kappa=x, y, z} \operatorname{Tr}\left(U \frac{\sigma_{\kappa}}{2} U^{\dagger} U_{\delta} \frac{\sigma_{\kappa}}{2} U^{\dagger}_{\delta}\right)
\end{equation}
in a two level system~\cite{Bow_02_PLA,Nie_02_PLA}, where $U$ is the target gate and
\begin{equation}\label{}
  U_{\delta}=\mathcal{T} \exp \left(-i \int_{0}^{t} H\left(\delta,t^{\prime}\right) d t^{\prime}\right)
\end{equation}
is the unitary evolution operator, with $\mathcal{T}$ being the time ordering operator.

Under inhomogeneous broadening, the objective function is then written as
\begin{equation}
  \label{aveFgate}
  F^G_\text{obj}=\mathcal{N}\sum_{k=1}^{M}p(\delta_k)f_g(\delta_k),
\end{equation}
where $\mathcal{N}$ and $p(\delta_k)$ take the same form as those in Eq.~\eqref{aveFd}. The optimized results of the PM ($N=1,2$) and SFB-P2 ($N=1,5$) methods with the Hadamard gate, Pauli-X, Pauli-Y and Pauli-Z gate are shown in Fig.~\ref{fig9}. The parameters used in the optimization process are $T=100$ ns, $W/2\pi=10$ MHz, and $\Omega_{\text{max}}/2\pi=10$ MHz. Compared with the SFB-P2, the PM method no longer shows absolute superiority (e.g., for the Pauli-Y gate), but still shows its utility in view of versatility for different gate types and resource economy.

The robustness of the optimized control field under detunings and field amplitude variations are further demonstrated in Fig.~\ref{fig10}. Compared with the rectangular pulse with the same maximal amplitude, optimized pulses given by the PM method shows a general better robustness, although the degree of improvement is less obvious compared with the state transition case in Fig.~\ref{fig-amp}(d).

Single qubit gates are the fundamental tools for dynamical decoupling (DD)~\cite{Vio_98_pra} and nuclear spin detection~\cite{Tam_12_prl,Kol_12_prl}. The most widely used DD technique, including the Carr-Purcell sequence~\cite{Car_54_PR} and the XY family sequence \cite{Gul_69_JourMR}, play important roles in eliminating the effects of environmental noise. In Fig.~\ref{fig11}, we give a simulation of the spin coherence time $T_2$ using one XY8 pulse sequence implemented using the PM optimal field and a rectangular $\pi$ pulse, respectively. The drift Hamiltonian of the system in the interaction picture is
\begin{equation}\label{H0_noise}
  H_{dr}(t)=\frac{\delta+\delta_d(t)}{2}\sigma_z,
\end{equation}
with a constant component of noise $\delta$ following a Gaussian distribution, and a dynamic Lorentzian noise $\delta_d(t)$, which is modeled as an Ornstein-Uhlenbeck process and approximated by~\cite{Gil_96_PRE}
\begin{equation}
\delta_d(t+\Delta t)=\delta_d(t) e^{-\Delta t/\tau}+\left[\frac{c \tau}{2}\left(1-e^{-2\Delta t/\tau}\right)\right]^{1 / 2} n,
\end{equation}
with $\tau$ and $c$ being the noise relaxation time and the diffusion constant, respectively, and $n$ representing a sample value of the unit normal distribution. We take noise parameters leading to a dephasing time of $T_2^*\approx20$ ns, that is $\tau=20\ \mu$s, a standard deviation $(c\tau/2)^{1/2}=2\pi\times 50$ KHz for the dynamical noise while $\delta$ follows a Gaussian distribution with a zero mean value and a FWHM of $2\pi\times26.5$ MHz for the static noise~\cite{Gen_20_prr}.

Due to the finite pulse length of the control field, the effect of noise cannot be neglected during the pulse, thus the total Hamiltonian in this period is
\begin{equation}\label{}
  H(t)=H_{dr}(t)+H_c(t),
\end{equation}
where $H_c(t)$ takes the form of Eq.~\eqref{Hp} for the PM control field and the constant form $(\pi/2T_\text{pulse})\sigma_{x(y)} $ for the rectangular Pauli-X (Y) pulse, with $T_\text{pulse}$ the pulse length. We use the phase-modulated method with $N=1$, $W/2\pi=26.5$ MHz, and $\Omega_{\text{max}}/2\pi=10$ MHz to optimize the Pauli-X and Pauli-Y gate in an XY-8 sequence with pulse length $T_\text{pulse}=100$ ns, and the pulse length of the rectangular pulse for the standard XY-8 sequence is set to be $T_\text{pulse}=50$ ns, such that the two kinds of pulses have the same maximal amplitude. The pulse interval is $\tau_\text{pulse} \in [0.4, 5.6]\ \mu$s for the PM pulses and $\tau_\text{pulse} \in [0.45, 5.65]\ \mu$s for the rectangular pulses, such that the total evolution time $T$ for both methods is $T\in [4, 45.6]\ \mu$s.

The simulated coherence time using the PM optimal fields is $T_2\approx 39\ \mu$s, thereby increasing the $T_2\approx 24\ \mu$s using rectangular pulses by roughly one half. We note that the SFB-P2 method with a similar number of parameters will likewise give good results in such simulations and is also worth investigating for further application, where the two methods could be combined in order to complement each other.  With further exploration and more target designs, the PM method, as well as the SFB-P2 method can be used to enhance the robustness of DD sequences in different scenarios and thus have a high potential to be applied in quantum sensing experiments for an improvement of the detection precision.

\section{Conclusion}\label{sec_con}
We have introduced a new direct-optimization method based on a phase-modulated (PM) function basis that is feasible for applications in robust quantum control problems. As an example, we apply this method to find robust control fields for the control of an ensemble of two-level systems exhibiting an  inhomogeneous broadening. Constrained by the same maximal field strength and the same maximal search resources, the PM method reaches highly improved results compared with the widely used standard Fourier basis (SFB) and comparable results with the phase-introduced standard Fourier basis (SFB-P2), using one order of magnitude less optimization time. A detailed analysis of the overall optimization results further reveals that the PM method shows a stable capability to find the global optimum in the parameter landscape with the comparable or lower search resources, while SFB-P2 with more parameters is unable to plausibly obtain the same results in every trial run with the same search resources. A significant advantage of our PM method is the involvement of multiple frequency components in a single element of the PM function basis. This is shown in the frequency spectrum of the control fields during the optimization process. Considering possible drifts of the control field amplitudes and decoherence noise, when applied in a realistic model of an ensemble system, the PM method also shows a significantly stronger robustness in both of these situations,
compared with the SFB-P2 method in the same low-dimensional parameter space. As a further example, we also demonstrate the utilization of the PM function basis in the optimization of robust gate control fields for dynamical decoupling, leading to a prolonged coherence time of the system compared to traditional rectangular pulses.

We note that while the results we present in this paper are produced using a direct search optimization solver based on the Nelder-Mead method, similar results can also be obtained using a gradient-based solver. This implies that the relative advantage of the PM method over SFB methods relies on the form of basis rather than the specific solver, and can thereby be utilized beyond a direct search optimization. For the specific problem we discussed in this paper, the use of randomization of some parameters is not recommended since it degrades the probability of finding the global optimum. However, applying the randomization of parameters to the PM method is nevertheless worth considering and might be useful in other contexts.

With short optimization times and a strong robustness, the proposed PM method provides a potentially superior gradient-free optimization method. It has a high potential value in the field of spin-ensemble manipulations and optimal control in many-body physics, where the expensive evaluation process of the objective functions makes the total optimization process time-consuming, as the parameter space expands quickly with the system size. Furthermore, our proposed method can be straightforwardly adapted and integrated to advanced and highly-developed optimization packages, e.g., Remote Dressed CRAB~(RedCRAB), which is capable of performing closed-loop optimization remotely~\cite{HeckE11231}, and has been recently used to experimentally demonstrate the generation of genuine multipartite entanglement in the form of 20-qubit Greenberger-Horne-Zeilinger~(GHZ) states with Rydberg atoms~\cite{Omran570}. In addition, the capabilities to conveniently optimize gate robustness and prolong the coherence time using XY-8 sequences make the PM method potentially useful for improving the performance of DD techniques under inhomogeneous broadening ~\cite{wang_20_Sym} and quantum sensing experiments~\cite{Gen_20_prr}.

\section*{Acknowledgments}
We thank Christian Arenz and Yaoming Chu for helpful discussions and comments. This work is supported by the National Natural Science Foundation of China (Grant No. 11804110, No. 11950410494, and No. 11874024) and the National Key R$\&$D Program of China (2018YFA0306600). R.S.S. and F.J. acknowledges support from ERC Synergy Grant HyperQ, the German Federal Ministry of Education and Research (BMBF), DFG, VW Stiftung and EU via project ASTERIQS).

\bibliography{PiteBib}
\end{document}